\documentclass[10pt,english,prd,twocolumn,superscriptaddress,nofootinbib,preprintnumbers,showpacs,floatfix]{revtex4-2}
\usepackage{amsmath,amssymb,amsthm,hyperref}
\usepackage[utf8]{inputenc}
\usepackage{graphicx}
\usepackage{xcolor}

\begin{document}

\title{Dyonic Ellis-Bronnikov wormholes from warped extra dimensions}

\author{Francisco S. N. Lobo}
\email{fslobo@ciencias.ulisboa.pt}
\affiliation{Instituto de Astrof\'{i}sica e Ci\^{e}ncias do Espa\c{c}o, Faculdade de Ci\^{e}ncias da Universidade de Lisboa, Edifício C8, Campo Grande, P-1749-016 Lisbon, Portugal}
\affiliation{Departamento de F\'{i}sica, Faculdade de Ci\^{e}ncias da Universidade de Lisboa, Edif\'{i}cio C8, Campo Grande, P-1749-016 Lisbon, Portugal}

\author{Miguel A. S. Pinto}
\email{mapinto@ciencias.ulisboa.pt}
\affiliation{Instituto de Astrof\'{i}sica e Ci\^{e}ncias do Espa\c{c}o, Faculdade de Ci\^{e}ncias da Universidade de Lisboa, Edifício C8, Campo Grande, P-1749-016 Lisbon, Portugal}
\affiliation{Departamento de F\'{i}sica, Faculdade de Ci\^{e}ncias da Universidade de Lisboa, Edif\'{i}cio C8, Campo Grande, P-1749-016 Lisbon, Portugal}
\affiliation{William H. Miller III Department of Physics and Astronomy, Johns Hopkins University, 3400 North Charles Street, Baltimore, Maryland, 21218, USA}

\author{Manuel E. Rodrigues} 
\email{esialg@gmail.com}
\affiliation{Faculdade de F\'{i}sica, Programa de P\'{o}s-Gradua\c{c}\~{a}o em F\'{i}sica, Universidade Federal do Par\'{a}, 66075-110, Bel\'{e}m, Par\'{a}, Brazill}
\affiliation{Faculdade de Ci\^{e}ncias Exatas e Tecnologia, Universidade Federal do Par\'{a}, Campus Universit\'{a}rio de Abaetetuba, 68440-000, Abaetetuba, Par\'{a}, Brazil}

\date{\today}

\begin{abstract}
We investigate traversable wormhole solutions within a four-dimensional effective theory derived from a five-dimensional Einstein-Maxwell-Chern-Simons action with a non-minimally coupled scalar field. A warped Kaluza-Klein compactification yields an Einstein-frame theory containing a phantom dilaton, a canonical axion, a Maxwell field, and a Kaluza-Klein vector, with the couplings fixed by the higher-dimensional origin.
Focusing on the Ellis-Bronnikov geometry, we construct solutions that incorporate both dyonic Maxwell and Kaluza-Klein fields. For exponential gauge couplings, the Einstein equations determine the scalar kinetic term and the combined potentials, while the remaining field equations reduce to algebraic relations fixing the individual potentials and the radial behaviour of the electric charges. We obtain a systematic classification of configurations, ranging from the pure phantom-supported wormhole to fully coupled dilaton-axion-gauge configurations.
The Kaluza-Klein sector enriches the solution space with additional structure while preserving analytic tractability. These results show that regular, asymptotically flat traversable four-dimensional wormholes arise naturally from higher-dimensional scalar-tensor theories.
\end{abstract}

\maketitle

{\it Introduction.}
Traversable wormholes are among the most intriguing solutions of general relativity, offering hypothetical shortcuts between distant spacetime regions~\cite{Morris:1988cz,Morris:1988tu}. Despite specific scenarios where exotic matter is not required to sustain wormholes \cite{Harko:2013yb,Blazquez-Salcedo:2020czn,Konoplya:2021hsm}, their typical realisation relies on matter violating the null energy condition, most naturally provided by phantom scalar fields. The Ellis-Bronnikov wormhole remains the prototypical example of a static, traversable geometry supported by a massless phantom scalar~\cite{Ellis:1973yv,Bronnikov:1973fh}. A central motivation in recent years has been to understand how such configurations can be embedded in more complete physical settings, particularly through the inclusion of additional fields and interactions that arise naturally in fundamental theories.

In this direction, electrically charged extensions have been constructed in Einstein--Maxwell--dilaton theory with a phantom dilaton~\cite{Goulart:2017iko}, as well as in Einstein--Maxwell--scalar models with non-minimal couplings~\cite{Huang:2019arj}. These studies show that gauge fields can significantly modify the wormhole geometry, sometimes inducing horizons and leading to black-bounce or regular black hole configurations. While such results highlight the richness of the solution space, they also raise the question of whether these matter sectors can be derived systematically, rather than introduced ad hoc, and how their interplay affects the existence of regular, horizonless geometries.

Higher-dimensional theories provide a natural framework to address these issues, as compactification generically produces scalar and vector fields with well-defined couplings. Warped extra dimensions have been explored in the construction of Lorentzian wormholes, and in attempts to soften energy condition violations~\cite{Kar:2022omn,Chen:2000yi,Simmonds:2025tet,Deshpande:2022zfm}, while higher-curvature models admit solutions with improved properties~\cite{Zangeneh:2014noa,Mehdizadeh:2015jra,Mehdizadeh:2015dta,KordZangeneh:2015dks,Mehdizadeh:2016nna}. The idea of partly phantom fields, where the exotic behaviour is localised, further refines this picture~\cite{Torii:2013xba,Bronnikov:2022bud}. Despite these advances, a systematic derivation of coupled phantom--axion--gauge wormhole solutions directly from higher-dimensional compactifications remains largely unexplored.

In this Letter, we address this gap by deriving a four-dimensional effective theory from a five-dimensional Einstein--Maxwell system~\cite{Hollands:2007qf} with a Chern--Simons term~\cite{Alexander:2009tp}, a scalar potential, and a non-minimal coupling to gravity. Warped compactification yields an Einstein-frame theory containing a phantom dilaton $\Psi$, a canonical axion $\varphi$~\cite{Svrcek:2006yi}, a Maxwell field $A_\mu$, and a Kaluza--Klein vector $\mathcal{A}_\mu$~\cite{Overduin:1997sri}. For static, spherically symmetric Ellis--Bronnikov backgrounds, we construct dyonic solutions with exponential couplings. The Einstein equations fix the scalar sector, while the remaining equations reduce to algebraic constraints determining the potentials and charges. We classify configurations from the pure phantom case to fully coupled dilaton--axion--gauge systems, showing that the higher-dimensional origin enriches the solution space while preserving analytic control and yielding regular, asymptotically flat wormholes.\\

{\it Five-dimensional action and compactification.}
Our starting point is a five‑dimensional theory, working in units where \(8\pi G_{(5)}=1\), whose action reads
\begin{eqnarray}\label{eq:5D}
	S_{(5)} &=& \int d^5x \sqrt{-g_{(5)}} \Big[ \tfrac12(1-\xi\Phi^2)R_{(5)} - \tfrac12(\partial\Phi)^2 - V(\Phi) \nonumber \\
	&& \qquad\qquad - \tfrac14 F_{MN}F^{MN} \Big] + \frac{\lambda}{8}\int A\wedge F\wedge F \, .
\end{eqnarray}
It contains a dilaton \(\Phi\) non-minimally coupled to gravity through the parameter \(\xi\) and a potential \(V(\Phi)\), a gauge field strength \(F_{MN}\), and a Chern–Simons term with coupling constant \(\lambda\).

We compactify the fifth dimension on a circle of circumference \(L\) using a warped Kaluza–Klein ansatz,
\begin{equation}
	ds_{(5)}^2 = e^{2\alpha\Phi} \tilde g_{\mu\nu}dx^\mu dx^\nu + e^{2\beta\Phi}\big(dw + \kappa\mathcal{A}_\mu dx^\mu\big)^2,
\end{equation}
where \(w\in[0,L)\) parametrises the compact direction. The constants \(\alpha,\beta\) control the dilaton warping, while \(\kappa\) fixes the normalization of the Kaluza–Klein vector \(\mathcal{A}_\mu\). The five‑dimensional gauge field decomposes as
\begin{equation}
	A_M dx^M = A_\mu dx^\mu + \varphi\,dw,
\end{equation}
giving rise to a four‑dimensional Maxwell field \(A_\mu\) and an axion \(\varphi\).

By integrating over the compact coordinate and performing a Weyl rescaling \(\tilde g_{\mu\nu} = \Omega^2 g_{\mu\nu}\) to reach the Einstein frame (a detailed derivation is provided in Appendix~\ref{sec:5Dto4D}), we obtain the four‑dimensional effective action
\begin{eqnarray}\label{eq:4D}
	S_{(4)} &=& \int d^4x\sqrt{-g}\Big[ \tfrac12 R + \tfrac12(\partial\Psi)^2 - \tfrac12(\partial\varphi)^2 \nonumber \\
	&& \quad - \tfrac14\mathcal{B}(\Psi)F_{\mu\nu}F^{\mu\nu} - \tfrac14\mathcal{C}(\Psi)\mathcal{F}_{\mu\nu}\mathcal{F}^{\mu\nu} \nonumber \\
	&& \quad - \mathcal{V}(\Psi) - \mathcal{U}(\varphi) \Big] - \frac{\lambda L}{16}\int \varphi\,F\wedge F \, .
\end{eqnarray}
Here, \(\Psi\) is the canonically normalized dilaton, related to \(\Phi\), while \(\mathcal{F}_{\mu\nu}=\partial_\mu\mathcal{A}_\nu-\partial_\nu\mathcal{A}_\mu\). The field‑dependent couplings inherit the higher‑dimensional parameters, while the axion potential \(\mathcal{U}(\varphi)\) remains free, to be specified by the phenomenological context. A crucial feature of this construction is that, for appropriate choices of the warp exponents \((\alpha,\beta)\) and the non‑minimal coupling \(\xi\), the dilaton \(\Psi\) turns into a phantom field, with kinetic term \(+\tfrac12(\partial\Psi)^2\), providing the exotic matter needed to sustain a wormhole throat. The axion and the gauge fields, by contrast, contribute ordinary, positive‑energy matter.\\

{\it Field equations and static wormhole ansatz.}
The equations of motion follow from varying the four-dimensional action \eqref{eq:4D}. The Einstein equations take the standard form
\begin{equation}
	G_{\mu\nu} = \sum_i T^{(i)}_{\mu\nu},
\end{equation}
where each matter sector contributes its own stress-energy tensor. Explicitly, one finds
\begin{subequations}
	\begin{align}
	T^{(\Psi)}_{\mu\nu} &= -\partial_\mu\Psi\,\partial_\nu\Psi + \tfrac12 g_{\mu\nu}(\partial\Psi)^2,\\
	T^{(\varphi)}_{\mu\nu} &= \partial_\mu\varphi\,\partial_\nu\varphi - \tfrac12 g_{\mu\nu}(\partial\varphi)^2,\\
	T^{(A)}_{\mu\nu} &= \mathcal{B}(\Psi)\Big(F_{\mu\rho}F_\nu{}^\rho - \tfrac14 g_{\mu\nu}F^2\Big),\\
	T^{(\mathcal{A})}_{\mu\nu} &= \mathcal{C}(\Psi)\Big(\mathcal{F}_{\mu\rho}\mathcal{F}_\nu{}^\rho - \tfrac14 g_{\mu\nu}\mathcal{F}^2\Big),\\
	T^{(\mathcal{V},\mathcal{U})}_{\mu\nu} &= -g_{\mu\nu}\big(\mathcal{V}(\Psi)+\mathcal{U}(\varphi)\big).
	\end{align}
\end{subequations}
These expressions clearly separate the kinetic contributions of the scalars, the gauge-field energies weighted by their dilaton-dependent couplings, and the potential terms acting as effective sources of vacuum energy.

The scalar fields obey
\begin{subequations}
	\begin{align}
	\Box\Psi + \mathcal{V}'(\Psi) + \tfrac14\mathcal{B}'(\Psi)F^2 + \tfrac14\mathcal{C}'(\Psi)\mathcal{F}^2 &= 0,\\
	\Box\varphi - \mathcal{U}'(\varphi) - \tfrac{\lambda L}{8}\,{}^*\!F^{\mu\nu}F_{\mu\nu} &= 0,
	\end{align}
\end{subequations}
while the gauge-field equations are
\begin{subequations}
	\begin{align}
	\nabla_\nu\big(\mathcal{B}(\Psi) F^{\mu\nu}\big) + \tfrac{\lambda L}{2}\,{}^*\!F^{\mu\nu}\partial_\nu\varphi &= 0,\\
	\nabla_\nu\big(\mathcal{C}(\Psi) \mathcal{F}^{\mu\nu}\big) &= 0.
		\label{eq:Cherns-Simon}
	\end{align}
\end{subequations}
The axion thus couples directly to the topological density $F\wedge F$, providing a source term that will play an important role in charged configurations.

\medskip

To explore wormhole geometries, we restrict to static, spherically symmetric spacetimes of the form~\cite{Morris:1988cz}: 
\begin{equation}
	ds^2 = -dt^2 + \frac{dr^2}{1-b(r)/r} + r^2 d\Omega^2,
\end{equation}
where $b(r)$ is the shape function. A wormhole throat is located at $r=a$, defined by $b(a)=a$, together with the flaring-out condition $b - r b' > 0$, which ensures that the geometry opens up rather than pinches off. We consider a zero redshift function, for simplicity.

Working in an orthonormal frame $(e^{\hat t},e^{\hat r},e^{\hat\theta},e^{\hat\phi})$ simplifies the interpretation of the field equations. The non-vanishing components of the Einstein tensor are
\begin{equation}
	G_{\hat t\hat t} = \frac{b'}{r^2}, \quad
	G_{\hat r\hat r} = -\frac{b}{r^3}, \quad
	G_{\hat\theta\hat\theta} = G_{\hat\phi\hat\phi} = \frac{b - r b'}{2r^3}.
\end{equation}
These directly encode the energy density and pressures required to sustain the wormhole.

For the Maxwell field, we consider a dyonic configuration carrying both electric $Q$ and magnetic $P$ charges,
\begin{equation}
	F_{\hat t\hat r} = \frac{Q}{r^2}, \qquad
	F_{\hat\theta\hat\phi} = \frac{P}{r^2}.
\end{equation}
In full generality, the Kaluza--Klein vector $\mathcal{A}_\mu$ may also be active, with an analogous dyonic ansatz
\begin{equation}
	\mathcal{F}_{\hat t\hat r} = \frac{Q_{\rm KK}}{r^2}, \qquad
	\mathcal{F}_{\hat\theta\hat\phi} = \frac{P_{\rm KK}}{r^2}.
\end{equation}
With these ansätze, the various matter contributions take a particularly transparent form:
\begin{subequations}
	\begin{align}
	T^{(\Psi)}_{\hat t\hat t} &= T^{(\Psi)}_{\hat r\hat r} = -\tfrac12(\Psi')^2\Big(1-\frac{b}{r}\Big),\\
	T^{(\varphi)}_{\hat t\hat t} &= T^{(\varphi)}_{\hat r\hat r} = \tfrac12(\varphi')^2\Big(1-\frac{b}{r}\Big),\\
	T^{(A)}_{\hat t\hat t} &= -T^{(A)}_{\hat r\hat r} = \frac{\mathcal{B}(\Psi)}{2r^4}(Q^2+P^2),\\
	T^{(\mathcal{A})}_{\hat t\hat t} &= -T^{(\mathcal{A})}_{\hat r\hat r} = \frac{\mathcal{C}(\Psi)}{2r^4}(Q_{\rm KK}^2+P_{\rm KK}^2),\\
	T^{(\mathcal{V},\mathcal{U})}_{\hat t\hat t} &= \mathcal{V}+\mathcal{U}, \qquad
	T^{(\mathcal{V},\mathcal{U})}_{\hat r\hat r} = -(\mathcal{V}+\mathcal{U}),
	\end{align}
\end{subequations}
with analogous expressions for the angular components.

The Kaluza-Klein vector introduces no qualitatively new features: its stress-energy tensor has the same Maxwell‑like structure, it carries its own dilaton coupling $\mathcal{B}(\Psi)$, and it contributes additively to the total energy budget. Including this sector therefore, enlarges the parameter space, i.e., adding the constants $Q_{\rm KK}, P_{\rm KK}$ and the coupling function $\mathcal{C}(\Psi)$, without altering the essential solvability of the field equations.  For many purposes, the KK field can be set to zero to simplify the analysis, but retaining it enriches the family of admissible wormhole solutions and provides a more complete picture of the matter content descending from the five‑dimensional theory.\\

{\it Reduced independent equations.}
To streamline the equations, we introduce effective quantities that capture the distinct physical contributions:
\begin{subequations}
	\begin{align}
	\mathcal{K} &:=  \tfrac12\big[-(\Psi')^2 + (\varphi')^2\big]\Big(1-\frac{b}{r}\Big),
	\label{aux1}\\
	\mathcal{E} &:=  \frac{\mathcal{B}(\Psi)}{2r^4}\,(Q^2+P^2) + \frac{\mathcal{C}(\Psi)}{2r^4}\,(Q_{\rm KK}^2+P_{\rm KK}^2),
	\label{aux2}\\
	\mathcal{P} &:=  \mathcal{V}(\Psi)+\mathcal{U}(\varphi).
	\label{aux3}
	\end{align}
\end{subequations}
Here, $\mathcal{K}$ represents the net kinetic contribution of the scalar sector (highlighting the competition between the dilaton and axion), $\mathcal{E}$ is the total energy density of both Abelian gauge fields (Maxwell and Kaluza--Klein), and $\mathcal{P}$ collects all potential terms.

In terms of these variables, the Einstein equations simplify considerably and reduce to two independent relations:
\begin{subequations}
	\begin{align}
		\frac{b'}{r^2} &= \mathcal{K} + \mathcal{E} + \mathcal{P},
		\label{eq:indep1_corr}
			\\
		\frac{b'}{r^2} + \frac{b}{r^3} &= 2\left(\mathcal{E} + \mathcal{P}\right).
		\label{eq:indep2_corr}
	\end{align}
\end{subequations}
These equations admit a transparent interpretation: the first determines how the shape function $b(r)$ is sourced by the total energy density, while the second encodes the balance between radial and angular pressures. By combining them, one can isolate the kinetic contribution, $\mathcal{K} = \left(b'r - b \right)/(2r^3)$.
Moreover, the flaring-out condition required for a wormhole throat, $b -r b' > 0$, translates into a simple constraint on the scalar sector, namely, $\mathcal{K} < 0$, which implies $(\Psi')^2 > (\varphi')^2$ in the vicinity of the throat.
Thus, the dilaton must dominate over the axion in order to provide the effective exotic matter supporting the geometry.  The inclusion of the Kaluza--Klein field does not alter this conclusion; it merely augments the ordinary matter content through the additional term in $\mathcal{E}$.

\medskip

The scalar field equations also take a compact radial form. For the dilaton, one finds
\begin{eqnarray}
	&&\frac{\sqrt{1-\frac{b}{r}}}{r^2}\frac{d}{dr}\left[r^2\sqrt{1-\frac{b}{r}}\;\Psi'\right]
	+ \mathcal{V}'(\Psi)
	+ \frac{\mathcal{B}'(\Psi)}{2r^4}(P^2 - Q^2)
			\nonumber \\
	&& \qquad \qquad \qquad \qquad + \frac{\mathcal{C}'(\Psi)}{2r^4}(P_{\rm KK}^2 - Q_{\rm KK}^2) = 0,
	\label{eq:scalar1_corr}
\end{eqnarray}
while the axion satisfies
\begin{equation}
	\frac{\sqrt{1-\frac{b}{r}}}{r^2}\frac{d}{dr}\left[r^2\sqrt{1-\frac{b}{r}}\;\varphi'\right]
	- \mathcal{U}'(\varphi)
	- \frac{\lambda L}{2}\frac{QP}{r^4} = 0.
	\label{eq:scalar2_corr}
\end{equation}
The dilaton equation gains a Kaluza-Klein contribution mirroring the Maxwell term with coupling $\mathcal{B}(\Psi)$ and charges $Q_{\rm KK},P_{\rm KK}$. The axion equation remains unchanged, as the KK field lacks a Chern-Simons coupling; only the Maxwell charges $Q,P$ source the axion through the topological term.

Finally, the gauge sector is constrained by the equations of motion. For the Maxwell field, one obtains
\begin{equation}
	\frac{d}{dr}\bigl( \mathcal{B}(\Psi)\,Q \bigr)
	- \frac{\lambda L}{2}\,P\,\varphi' = 0.
	\label{eq:Maxwell_static_constraint}
\end{equation}
For the Kaluza-Klein vector, the absence of a Chern-Simons term leads to the simpler relation
\begin{equation}
	\frac{d}{dr}\bigl( \mathcal{C}(\Psi)\,Q_{\rm KK} \bigr) = 0,
	\qquad
	P_{\rm KK} = \text{const}.
	\label{eq:KK_static_constraint}
\end{equation}
Thus, the KK effective electric charge $\mathcal{C}(\Psi) Q_{\rm KK}$ is strictly conserved, while the magnetic charge $P_{\rm KK}$ is constant by virtue of the Bianchi identity.

Together, these form a closed system for the radial metric and matter profiles, explicitly intertwining geometry, scalars, and the two Abelian gauge fields in the wormhole construction.\\

{\it Ellis-Bronnikov wormhole}.
We illustrate this construction by embedding both a dyonic Maxwell field and a dyonic Kaluza--Klein vector into the classic Ellis--Bronnikov wormhole. Taking the geometry as a starting point, we determine the matter content required to consistently sustain it within the full set of field equations, thereby revealing how gauge and scalar sectors combine to support the wormhole configuration.

The Ellis-Bronnikov shape function is given by $b(r) = a^2/r$,
which describes the spatial geometry of the Ellis--Bronnikov wormhole.  The throat is located at \(r=a\) and the flaring‑out condition \(b-rb'>0\) holds identically.

We consider the general matter content of our model: a phantom dilaton \(\Psi\), a canonical axion \(\varphi\), a dyonic Maxwell field carrying charges \(Q\) and \(P\), and a dyonic Kaluza-Klein vector field carrying charges \(Q_{\rm KK}\) and \(P_{\rm KK}\).

In low-energy effective string models whose target space includes a four-dimensional manifold \(M_4\) and a two-dimensional hyperbolic moduli space, the dilaton and axion unify into a complex scalar \(S\) that parametrises that curved internal geometry. This structure survives dimensional reduction down to four dimensions, where the axio-dilaton sector becomes a nonlinear sigma model, minimally coupled to gravity~\cite{Bakas:1996dz}. Because the action~\eqref{eq:4D} exhibits such minimal coupling, under static spherical symmetry, the equations of motion for the dilaton and axion can be combined into a single complex equation. Configurations that preserve the residual phase-rotation symmetry of the internal space automatically align the radial gradients of the real and imaginary parts of \(S\), leading to the proportionality ansatz
\begin{equation}
	\varphi'(r) = c\,\Psi'(r), \qquad |c|<1, \label{ansatz}
\end{equation}
where \(c\) encodes the constant phase of the complex modulus. The bound \(|c|<1\) guarantees that the dilaton retains its phantom character, which is essential for sustaining a traversable wormhole.

The effective electric charges are \(q(r) \equiv \mathcal{B}(\Psi) Q(r)\) and \(q_{\rm KK}(r) \equiv \mathcal{C}(\Psi) Q_{\rm KK}(r)\), while the magnetic charges \(P\) and \(P_{\rm KK}\) are strictly constant by virtue of the Bianchi identities.  To proceed, we adopt exponential coupling functions typical of dimensionally reduced theories,
\begin{equation}
	\mathcal{B}(\Psi) = \mathcal{B}_0\,e^{\beta\Psi}, \quad
	\mathcal{C}(\Psi) = \mathcal{C}_0\,e^{\gamma\Psi}, \quad \mathcal{B}_0,\mathcal{C}_0>0.
\end{equation}
Thus, using the auxiliary quantities \eqref{aux1}--\eqref{aux3}, the independent Einstein equations\eqref{eq:indep1_corr}--\eqref{eq:indep2_corr} provide
\begin{subequations}
	\begin{align}
		(1-c^2)(\Psi')^2\left(1-\frac{a^2}{r^2}\right) =& \frac{2a^2}{r^4}, \label{eq:EBgen_kinetic_dyonKK}\\
		\mathcal{V}+\mathcal{U} =& -\frac{\mathcal{B}_0 e^{\beta\Psi}}{2r^4}(Q^2+P^2) 
			\nonumber \\
		& \hspace{-0.75cm} - \frac{\mathcal{C}_0 e^{\gamma\Psi}}{2r^4}(Q_{\rm KK}^2+P_{\rm KK}^2). \label{eq:EBgen_VU_dyonKK}
	\end{align}
\end{subequations}
Equation \eqref{eq:EBgen_kinetic_dyonKK} determines the dilaton gradient independently of the potentials and the gauge charges.  

\begin{figure}[!h]
\begin{center}
\begin{tabular}{ccc}
\includegraphics[width=\columnwidth]{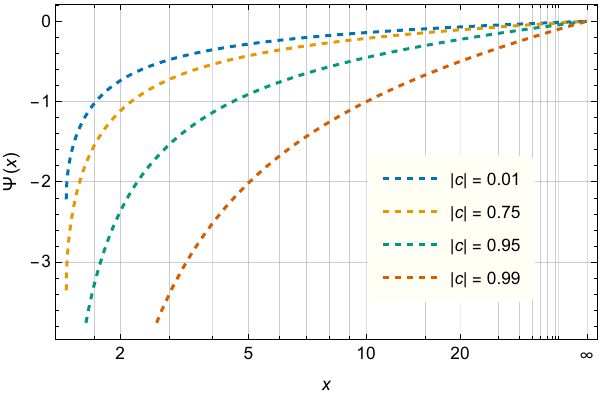}
\end{tabular}
\end{center}
\caption{Dilaton field $\Psi$ as a function of the adimensional coordinate $x=r/a$, for four different values of $|c|$. The coordinate $x$ ranges from $1$ to $\infty$, allowing the verification of the overall scaling of the dilaton profile independently of the throat parameter $a$.}
\label{plot-dilaton}
\end{figure}

Integrating the dilaton equation with the asymptotic condition $\Psi(\infty)=0$ gives
\begin{equation}
	\Psi(r) = -\frac{2}{\sqrt{2(1-c^{2})}}\;\operatorname{arccosec}\!\left(\frac{r}{a}\right),
	\label{eq:Psi_profile}
\end{equation}
while the axion follows from $\varphi'(r)=c\,\Psi'(r)$ and $\varphi(\infty)=0$ as
$\varphi(r)=c\,\Psi(r)$.  Thus, the axion is a rescaled dilaton, and the parameter $c$ controls the scalar mixing: as $c\to 1$, the two fields contribute equally to sustaining the wormhole geometry.
In Fig.~\ref{plot-dilaton}, the dilaton $\Psi$ is shown for different values of $|c|$. The profile remains regular across the coordinate $x$ and becomes progressively less steep as $|c|$ increases.

An important simplification occurs for the Ellis-Bronnikov metric together with the proportionality ansatz \eqref{ansatz}.  By massaging Eq. \eqref{eq:EBgen_kinetic_dyonKK}, one verifies that \(\mathcal{D}(r)\Psi'\), with \(\mathcal{D}(r)=r^2\sqrt{1-a^2/r^2}\), evaluates to a constant.  Consequently, \(\mathcal{D}(r)\varphi' = c\,\mathcal{D}(r)\Psi'\) is also constant. Again, this feature is \emph{not} generic; it is a special case of the \(b(r)=a^2/r\) geometry and the proportionality ansatz, but it drastically simplifies the scalar field equations, reducing them to purely algebraic consistency conditions. Therefore, the scalar field equations \eqref{eq:scalar1_corr}--\eqref{eq:scalar2_corr} become
\begin{subequations}
	\begin{align}
		\mathcal{V}'(\Psi) + \frac{\beta\mathcal{B}_0 e^{\beta\Psi}}{2r^4}(P^2 - Q^2) &
			\nonumber\\
		+ \frac{\gamma\mathcal{C}_0 e^{\gamma\Psi}}{2r^4}(P_{\rm KK}^2 - Q_{\rm KK}^2) &= 0, 
		\label{eq:EBgen_Psi_consistency_dyonKK}\\
		\mathcal{U}'(\varphi) + \frac{\lambda L}{2}\frac{QP}{r^4} &= 0. \label{eq:EBgen_varphi_consistency_dyonKK}
	\end{align}
\end{subequations}
Note that the Kaluza--Klein charges appear only in the dilaton equation, not in the axion equation, because the axion couples exclusively to the Maxwell sector via the Chern-Simons term.

The gauge sectors are constrained by their respective equations of motion. Using \(\varphi' = c\,\Psi'\), for the Maxwell field, Eq.~\eqref{eq:Maxwell_static_constraint} integates to yield
\begin{equation}
	Q(\Psi) = e^{-\beta\Psi} \left( Q_0 + \frac{\lambda L c}{2\mathcal{B}_0} P \, \Psi \right),
	\label{eq:Q_solution_KK}
\end{equation}
where \(Q_0\) is an integration constant. For the Kaluza--Klein vector, Eq.~\eqref{eq:KK_static_constraint} gives simply
\begin{equation}
	Q_{\rm KK}(\Psi) = Q_{\rm KK}^{(0)} e^{-\gamma\Psi},
	\label{eq:QKK_solution}
\end{equation}
with constant \(Q_{\rm KK}^{(0)}\).  The magnetic charges \(P\) and \(P_{\rm KK}\) remain constant.

With \(Q(\Psi)\) and \(Q_{\rm KK}(\Psi)\) known explicitly, Eq.~\eqref{eq:EBgen_VU_dyonKK} fixes the sum of the potentials,
\begin{equation}
	\mathcal{V} + \mathcal{U} = -\frac{\mathcal{B}_0 e^{\beta\Psi}}{2r^4}\bigl( Q(\Psi)^2 + P^2 \bigr) - \frac{\mathcal{C}_0 e^{\gamma\Psi}}{2r^4}\bigl( Q_{\rm KK}(\Psi)^2 + P_{\rm KK}^2 \bigr),
\end{equation}
while the consistency conditions \eqref{eq:EBgen_Psi_consistency_dyonKK} and \eqref{eq:EBgen_varphi_consistency_dyonKK} determine their individual derivatives.  Using the chain rule \(\mathcal{U}' = \frac{d\mathcal{U}}{d\varphi} = \frac{1}{c} \frac{d\mathcal{U}}{d\Psi}\), the second condition becomes
\begin{equation}
	\frac{d\mathcal{U}}{d\Psi} + \frac{\lambda L c}{2} \frac{Q(\Psi) P}{r^4} = 0.
\end{equation}
Together with the derivative of \eqref{eq:EBgen_VU_dyonKK}, this system can be integrated parametrically to obtain \(\mathcal{V}(\Psi)\) and \(\mathcal{U}(\Psi)\) (and hence \(\mathcal{U}(\varphi)\)).  The additional Kaluza--Klein terms contribute additively to the potential sum and to the dilaton consistency condition, but they do not alter the structural solvability of the equations.  As before, the explicit expressions involve integrals that are well‑defined and can be evaluated numerically for any chosen parameters.\\

The general parametric solution described above encompasses a broad family of wormhole configurations.  Several physically distinct limits are worth highlighting separately.

\paragraph{Pure phantom dilaton.}
Set \(c=0\), all charges to zero, \(\mathcal{U}=0\).  Then \(\mathcal{V}=0\) and the scalar profile reduces to \(\Psi(r) = \Psi_0 + \sqrt{2}\,\mathrm{arcsec}(r/a)\), the classic massless phantom wormhole.

\paragraph{Phantom dilaton + canonical axion.}
Set all charges to zero but keep \(c\neq0\) with \(|c|<1\).  Then \(\mathcal{V}+\mathcal{U}=0\); choosing \(\mathcal{V}=\mathcal{U}=0\) yields the profiles \eqref{eq:Psi_profile} with the amplitude enhanced by \((1-c^2)^{-1/2}\).

\paragraph{Phantom dilaton + dyonic Maxwell and Kaluza--Klein fields (no axion).}
Set \(c=0\), \(\mathcal{U}=0\), and allow the charges to be non‑zero.  The dilaton profile reverts to the pure phantom case.  Equations \eqref{eq:Q_solution_KK} and \eqref{eq:QKK_solution} give \(Q = Q_0 e^{-\beta\Psi}\) and \(Q_{\rm KK} = Q_{\rm KK}^{(0)} e^{-\gamma\Psi}\) (constant effective charges).  The potential is determined by \eqref{eq:EBgen_VU_dyonKK}, and the consistency condition \eqref{eq:EBgen_Psi_consistency_dyonKK} is automatically satisfied.

\paragraph{Full system: dilaton + axion + dyonic Maxwell + Kaluza--Klein.}
Keep all parameters non‑zero (\(|c|<1\), charges non‑zero).  The scalar profiles are \eqref{eq:Psi_profile}, the electric charges evolve according to \eqref{eq:Q_solution_KK} and \eqref{eq:QKK_solution}, and the potentials are obtained parametrically.  The Kaluza--Klein sector adds new independent constants and coupling exponents (\(\gamma\), \(\mathcal{C}_0\), \(Q_{\rm KK}^{(0)}\), \(P_{\rm KK}\)), thereby enriching the solution space without complicating the analytic structure.

For the Ellis-Bronnikov geometry, the scalar profiles are universal, fixed by the metric and \(c\), while gauge charges and potentials adjust to satisfy the remaining equations. The inverse method thus efficiently constructs regular, asymptotically flat wormholes with a rich dyonic content including both Maxwell and Kaluza-Klein fields.\\

{\it Conclusion.}
We have shown that a five-dimensional Einstein--Maxwell--Chern--Simons theory compactified on a warped circle yields a four-dimensional effective theory containing a phantom dilaton, a canonical axion, and Abelian gauge fields. Within this framework, regular, asymptotically flat traversable wormholes emerge in a systematic manner. Focusing on the Ellis--Bronnikov geometry, we consistently embed both dyonic Maxwell and Kaluza--Klein vector fields.
A key feature of the construction is the universality of the scalar sector: the scalar profiles are fully fixed by the geometry and axion--dilaton structure, independent of the gauge fields, while the gauge charges adjust dynamically to satisfy the remaining equations. This leads to a classification of solutions ranging from the pure phantom wormhole to fully coupled dilaton--axion--gauge configurations.

Adopting exponential gauge couplings motivated by dimensional reduction, we found that the electric charges evolve radially in response to the axion gradient via the Chern--Simons term, and the required scalar potentials are obtained parametrically.  The inclusion of the Kaluza--Klein sector enriches the solution space with additional parameters and coupling functions without compromising analytic solvability. These results highlight the power of the inverse method to generate exact, self‑consistent wormhole solutions with a rich matter content descending directly from higher‑dimensional physics.  Extensions to rotating geometries, stability analyses, and embeddings in explicit string compactifications offer promising directions for future work.\\

{\it Acknowledgments:}
FSNL and MASP acknowledge funding from the Funda\c{c}\~{a}o para a Ci\^{e}ncia e a Tecnologia (FCT) grant UID/04434/2025.
FSNL acknowledges support from the FCT Scientific Employment Stimulus contract with reference CEECINST/00032/2018. MASP acknowledges support from the FCT through Fellowship UI/BD/154479/2022. MAPS also acknowledged financial support from the Luso-American Development Foundation (FLAD) under an R\&D Internship Grant, from a Fulbright Grant for Research, supported by FCT, and support from the Spanish Grants PID2020-116567GB-C21, PID2023-149560NB-C21, funded by MCIN/AEI/10.13039/ 501100011033.
MER thanks Conselho Nacional de Desenvolvimento Cient\'{\i}fico e Tecnol\'{o}gico - CNPq, Brazil, for partial financial support.



\appendix

\section{Five-dimensional origin and dimensional reduction}
\label{sec:5Dto4D}

We start from a five‑dimensional theory, working in units where the five‑dimensional Newton constant is set to \(8\pi G_{(5)}=1\). The complete action reads
\begin{eqnarray}\label{eq:5Dapp}
	S_{(5)} &=& \int d^5x \sqrt{-g_{(5)}} \Big[ \tfrac12(1-\xi\Phi^2)R_{(5)} - \tfrac12(\partial\Phi)^2 - V(\Phi) \nonumber \\
	&& \qquad\quad - \tfrac14 F_{MN}F^{MN} \Big] + \frac{\lambda}{8}\int A\wedge F\wedge F \, .
\end{eqnarray}
The model contains a scalar field \(\Phi\) (the dilaton) non‑minimally coupled to gravity through the parameter \(\xi\), with a self‑interaction potential \(V(\Phi)\).  The gauge sector is described by the field strength \(F_{MN}=\partial_M A_N-\partial_N A_M\) and a Chern–Simons term with coupling \(\lambda\), a natural ingredient in higher‑dimensional supergravity and string‑inspired models.

\subsection{Warped Kaluza--Klein ansatz and dimensional reduction}

To extract the four‑dimensional effective theory, we compactify the fifth dimension on a circle of circumference \(L\).  We adopt a warped Kaluza--Klein ansatz for the five‑dimensional metric,
\begin{equation}
	ds_{(5)}^2 = e^{2\alpha\Phi} \tilde g_{\mu\nu}dx^\mu dx^\nu + e^{2\beta\Phi}\big(dw + \kappa\mathcal{A}_\mu dx^\mu\big)^2,
	\label{eq:KK_ansatz}
\end{equation}
where \(w\in[0,L)\) parametrises the compact direction.  The constants \(\alpha\) and \(\beta\) govern the dilaton warping, while \(\kappa\) sets the normalisation of the Kaluza--Klein vector field \(\mathcal{A}_\mu\) that emerges from the off‑diagonal metric components.  The four‑dimensional metric \(\tilde g_{\mu\nu}\) is initially in the Jordan frame, as the non‑minimal coupling \(\xi\Phi^2 R_{(5)}\) still multiplies the Ricci scalar.

The five‑dimensional gauge field is decomposed as
\begin{equation}
	A_M dx^M = A_\mu dx^\mu + \varphi\,dw,
	\label{eq:A_decomp}
\end{equation}
where \(A_\mu\) becomes a four‑dimensional Maxwell field and \(\varphi\) is a scalar (the axion) arising from the internal component.

Inserting the ansätze \eqref{eq:KK_ansatz} and \eqref{eq:A_decomp} into the five‑dimensional action \eqref{eq:5Dapp} and integrating over the compact coordinate \(w\) yields a four‑dimensional action in the Jordan frame.  
The five-dimensional Ricci scalar \(R_{(5)}\) computed from the warped metric decomposes into four-dimensional quantities as
\begin{eqnarray}
		&&R_{(5)} = e^{-2\alpha\Phi} \Big[ \tilde{R} - 2(3\alpha+\beta)\tilde{\Box}\Phi - \bigl(6\alpha^2 + 4\alpha\beta + \beta^2\bigr) \times
			\nonumber \\ 
		&& \qquad
		\times \tilde{g}^{\mu\nu}\partial_\mu\Phi\,\partial_\nu\Phi \Big] - \frac{\kappa^2}{4}\, e^{2\beta\Phi}\, \tilde{g}^{\mu\rho}\tilde{g}^{\nu\sigma} \mathcal{F}_{\mu\nu}\mathcal{F}_{\rho\sigma} 
			\nonumber \\
		&& \quad + \text{total derivatives involving } \tilde{\nabla}_\mu\mathcal{A}^\mu \text{ and } \tilde{\nabla}_\mu\mathcal{F}^{\mu\nu},
	\label{eq:R5_full}
\end{eqnarray}
where \(\tilde{R}\) and \(\tilde{\Box}\) are the Ricci scalar and d'Alembertian of the four‑dimensional Jordan‑frame metric \(\tilde{g}_{\mu\nu}\), and \(\mathcal{F}_{\mu\nu} = \partial_\mu\mathcal{A}_\nu - \partial_\nu\mathcal{A}_\mu\) is the Kaluza–Klein field strength.
The gauge kinetic term gives
\begin{eqnarray}
	&&	-\frac14 \int d^5x \sqrt{-g_{(5)}} F_{MN}F^{MN} = 
	-\frac14 \int d^4x \sqrt{-\tilde g}\, L \times
	\nonumber \\
	&& \qquad \qquad \qquad \times \Big[ e^{\beta\Phi} F_{\mu\nu}F^{\mu\nu} + 2 e^{-\beta\Phi} (\partial\varphi)^2 \Big],
\end{eqnarray}
while the Chern–Simons term reduces to a topological coupling \(\propto \int \varphi\, F\wedge F\) in four dimensions.

After integrating by parts to remove the \(\Box\Phi\) term (which contributes a total derivative in the action), the four‑dimensional Jordan‑frame action takes the form
\begin{eqnarray}
	S_{(4)}^{\rm J} &=& \int d^4x \sqrt{-\tilde g} \Big[ \tfrac12 \tilde f(\Phi) \tilde R - \tfrac12 \tilde h(\Phi) (\partial\Phi)^2 
	- \tilde V(\Phi) \nonumber \\
	&&  - \tfrac14 \tilde{\mathcal{B}}(\Phi) F_{\mu\nu}F^{\mu\nu} - \tfrac14 \tilde{\mathcal{C}}(\Phi) \tilde{\mathcal{F}}_{\mu\nu}\tilde{\mathcal{F}}^{\mu\nu} 
	- \tfrac12 \tilde{\mathcal{D}}(\Phi) (\partial\varphi)^2 \Big] \nonumber \\
	&&  + \frac{\lambda L}{8} \int \varphi \, F\wedge F,
\end{eqnarray}
where the coupling functions are given explicitly by
\begin{eqnarray}
	\tilde f(\Phi) &=& L\, e^{2\alpha\Phi} (1-\xi\Phi^2), \nonumber\\
	\tilde h(\Phi) &=& L\, e^{2\alpha\Phi} \Big[ 1 - 6\alpha^2 - 4\alpha\beta - \beta^2 
	\nonumber \\
	&& \qquad \qquad
	+ 3\left(2\alpha +\beta
	- \frac{2\xi\Phi}{1-\xi\Phi^2}\right)^2 \Bigg], 
		\nonumber\\
	\tilde V(\Phi) &=& L\, e^{4\alpha\Phi} V(\Phi), 
	\qquad \tilde{\mathcal{B}}(\Phi) = L\, e^{\beta\Phi},
	\nonumber\\
	\tilde{\mathcal{C}}(\Phi) &=& \frac{L\kappa^2}{4} e^{2(\beta-\alpha)\Phi}, \qquad
	\tilde{\mathcal{D}}(\Phi) = L\, e^{-\beta\Phi}.
	\nonumber
\end{eqnarray}

\subsection{Weyl rescaling to the Einstein frame}

To remove the field‑dependent gravitational coupling \(\tilde f(\Phi)\) and bring the action to the canonical Einstein–Hilbert form, we perform a Weyl rescaling of the metric,
\begin{equation}
	\tilde g_{\mu\nu} = \Omega^2 g_{\mu\nu}, \quad \Omega^2 = \tilde f(\Phi)^{-1} = \frac{1}{L\, e^{2\alpha\Phi}(1-\xi\Phi^2)}.
	\label{eq:Weyl}
\end{equation}
Under this transformation, the Ricci scalar transforms as
\begin{equation}
	\tilde R = \Omega^{-2} \Big[ R - 6 g^{\mu\nu}\nabla_\mu\nabla_\nu\ln\Omega - 6 g^{\mu\nu}(\partial_\mu\ln\Omega)(\partial_\nu\ln\Omega) \Big].
\end{equation}
The kinetic term of the dilaton receives an additional contribution from the derivatives of \(\Omega\), leading to a new effective kinetic function.  After a total derivative cancellation, the scalar kinetic term becomes
\begin{equation}
	S_{\rm kin} = \int d^4x \sqrt{-g} \Big[ -\frac12 G_{\Phi\Phi} (\partial\Phi)^2 - \frac12 \tilde{\mathcal{D}}(\Phi) \Omega^{-2} (\partial\varphi)^2 \Big],
\end{equation}
with
\begin{equation}
	G_{\Phi\Phi} = 1 - 6\alpha^2 - 4\alpha\beta - \beta^2 + 3\Bigl(2\alpha+\beta - \frac{2\xi\Phi}{1-\xi\Phi^2}\Bigr)^2.
	\label{eq:GPhiPhi}
\end{equation}
The potentials rescale as \(\tilde V \to \Omega^{-4} \tilde V\), and the gauge kinetic terms as \(F_{\mu\nu}F^{\mu\nu} \to \Omega^{-4} F_{\mu\nu}F^{\mu\nu}\).

Finally, we canonically normalise the dilaton.  Depending on the choice of parameters \((\alpha,\beta,\xi)\), the function \(G_{\Phi\Phi}\) can be either positive or negative.  For our purposes we select the regime where \(G_{\Phi\Phi} < 0\); we then define the canonical phantom field 
\begin{equation}
	d\Psi = \sqrt{-G_{\Phi\Phi}}\, d\Phi.
	\label{eq:canonical}
\end{equation}\\
The kinetic term becomes \(+\frac12 (\partial\Psi)^2\), signalling a phantom nature.  The axion \(\varphi\) already possesses a standard canonical kinetic term \(-\frac12 (\partial\varphi)^2\) after absorbing the constant factor \(L e^{-\beta\Phi}\) into a trivial field redefinition (which we omit for simplicity, as \(\Phi\) is a slowly varying function along the wormhole).

\subsection{Four‑dimensional Einstein‑frame action}

Putting everything together, the four‑dimensional effective action in the Einstein frame reads
\begin{eqnarray}\label{eq:4Dapp}
	S_{(4)} &=& \int d^4x\sqrt{-g}\Big[ \tfrac12 R + \tfrac12(\partial\Psi)^2 - \tfrac12(\partial\varphi)^2 \nonumber \\
		&& \quad - \tfrac14\mathcal{B}(\Psi)F_{\mu\nu}F^{\mu\nu} - \tfrac14\mathcal{C}(\Psi)\mathcal{F}_{\mu\nu}\mathcal{F}^{\mu\nu} \nonumber \\
		&& \quad - \mathcal{V}(\Psi) - \mathcal{U}(\varphi) \Big] - \frac{\lambda L}{16}\int \varphi\,F\wedge F \, .
	\end{eqnarray}
	The field‑dependent couplings are inherited from the compactification:
	\begin{eqnarray}
		\mathcal{B}(\Psi) &=& L\, e^{\beta\Phi(\Psi)}, 
		\\[2pt]
		\mathcal{C}(\Psi) &=& \frac{L\kappa^2}{4}\, e^{2(\beta-\alpha)\Phi(\Psi)}\bigl[1-\xi\Phi(\Psi)^2\bigr], \\[2pt]
		\mathcal{V}(\Psi) &=& L\, e^{4\alpha\Phi(\Psi)} V\bigl(\Phi(\Psi)\bigr).
		\label{eq:couplings}
	\end{eqnarray}
	The axion potential \(\mathcal{U}(\varphi)\) is not fixed by the reduction and can be chosen according to the desired phenomenological model.
	
	An essential feature of this construction is the controlled emergence of a phantom sector: for appropriate warping exponents \(\alpha,\beta\) and non‑minimal coupling \(\xi\), the dilaton \(\Psi\) acquires a phantom kinetic term, providing the exotic matter necessary to sustain traversable wormhole throats, while the axion \(\varphi\) and the gauge fields contribute ordinary, positive‑energy matter.

\end{document}